\documentclass[acmsmall]{acmart}
\AtBeginDocument{%
  }

\setcopyright{cc}
\setcctype{by}
\acmJournal{PACMHCI}
\acmYear{2027} \acmVolume{11} \acmNumber{1} \acmArticle{GROUP18}
\acmMonth{1} \acmDOI{10.1145/3859404}
\usepackage{multirow} 
\usepackage{xcolor}
\usepackage{booktabs}
\usepackage{float}

\definecolor{blue}{HTML}{2D9CDB}
\definecolor{green}{HTML}{06BA49}
\definecolor{yellow}{HTML}{F79600}

\received{31 March 2026}
\received[revised]{17 July 2026}
\received[accepted]{17 September 2026}

\begin{document}

\title[Socially Acceptable 3D Feedback Design]{Exploring Socially Acceptable 3D Feedback Design for Inclusive Small-Group Conversation}

\author{Qianjie Wei}
\affiliation{%
  \department{Department of Computer Science}
  \institution{University of Rochester}
  \city{Rochester}
  \state{New York}
  \country{USA}}
\email{qwei3@ur.rochester.edu}
\orcid{0009-0001-4429-4499}

\author{Yifan Li}
\affiliation{%
  \department{Department of Computer Science}
  \institution{University of Rochester}
  \city{Rochester}
  \state{New York}
  \country{USA}}
\email{yli287@ur.rochester.edu}
\orcid{0000-0001-7887-3598}

\author{Zhen Bai}
\authornote{Corresponding author}
\affiliation{%
  \department{Department of Computer Science}
  \institution{University of Rochester}
  \city{Rochester}
  \state{New York}
  \country{USA}}
\email{zhen.bai@rochester.edu}
\orcid{0000-0002-3258-0228}

\renewcommand{\shortauthors}{Qianjie Wei, Yifan Li, and Zhen Bai}

\begin{abstract}
Prior research has shown that providing feedback on group dynamics can improve the quality of small-group conversations. However, publicly delivered feedback that targets specific individuals may reduce social acceptability due to negative experiences such as social pressure, embarrassment, and resistance. This work investigates how public feedback can be designed to remain socially acceptable while maintaining its effectiveness in XR-based small-group conversations. Informed by social science and HCI literature, we identified three design strategies (incorporate humor, reduce social attention, and use indirectness) and proposed a set of metaphor- and agent-based 3D feedback probes. We then conducted a two-phase study: (1) an online survey with 50 participants to examine how such feedback designs are perceived under public and private delivery, and (2) follow-up interviews with 14 participants who completed the survey to better understand their interpretations and elicit additional insights. Results show that metaphor- and agent-based feedback can reduce social discomfort while maintaining effectiveness. Interview findings further highlighted the role of three strategies in making feedback easier to accept, while requiring careful balance. We also discussed design implications for creating socially acceptable 3D feedback to support inclusive small-group conversations in XR.
\end{abstract}




\maketitle

\section{Introduction}

Imagine you are in a conversation with two other peers. One person keeps talking and no one else can get a word in. Or maybe you notice that someone is upset, but don’t know how to bring them back into the conversation. In moments like these, a feedback \cite{ruesch1956nonverbal} delivered in front of everyone may help the conversation get back on track effectively. But it creates a dilemma: the feedback may also single someone out, making them feel exposed, judged, or embarrassed.

Public feedback, i.e., feedback shown to everyone in the group \cite{ouverson2021analysis}, can be especially powerful in group settings: it can increase user motivation \cite{geister2006effects,mumm2011designing}, improve team awareness \cite{geister2006effects}, and address problematic behavior effectively due to social comparison \cite{dijkstra2008social} and social influence \cite{argyle1957social}. However, delivering feedback that targets individuals in front of the entire group may lead to negative psychological feelings, such as embarrassment or even resistance \cite{keltner1997embarrassment,archibald1971self}, making public feedback less acceptable. Private feedback, defined as feedback shown only to the person to whom it applies \cite{ouverson2021analysis}, can avoid this psychological discomfort \cite{houtti2025observe,morris2006mediating}, but it is often easy to be ignored and not effective \cite{argyle1957social,ohtsuki2015reputation}.

A substantial body of research in HCI and CSCW has investigated on providing feedback about group dynamics and supporting inclusive group conversations \cite{samrose2021meetingcoach,houtti2025observe,kim2020bot,schiavo2014overt,leshed2009visualizing, armstrong2024sealmates}, both in co-located and distributed settings. Existing feedback systems such as CoCo \cite{samrose2018coco}, ``Observe, Ask, Intervene'' \cite{houtti2025observe}, MeetingCoach \cite{samrose2021meetingcoach}, and SealMates \cite{armstrong2024sealmates} demonstrate that visualized conversational cues and behavioral patterns (e.g., speaking time, turn-taking, emotional tone) has a positive effect on the conversation. There are also emerging research of extended reality (XR), which encompasses virtual reality (VR), augmented reality (AR), and mixed reality (MR), in supporting group conversational dynamics such as 
turn-taking \cite{lee2024may,li2022conversation}
and balanced participation \cite{yang2023understanding}. 
While research shows that users in XR-mediated group interaction also experience social discomfort associated with public feedback \cite{slater2000small}, there is a gap of research on exploring design space of public feedback in XR environments to reduce such discomfort.

XR-based communication provides more possibilities for the use of digital representations (e.g., agents) \cite{liu2023boosting} and contextual cues (e.g., embodied metaphors) \cite{liu2023boosting,wang2018break}, which demonstrates the potential to offer novel feedback forms in group conversation.
Inspired by this, we proposed our research question (RQ): \textbf{How could we design socially acceptable and effective public feedback for supporting inclusive small-group conversations in XR environments?}

To address this RQ, we first ground our design in theories and findings from social science and HCI literature, and synthesize three strategies for reducing embarrassment in public settings, including \textbf{incorporating humor} \cite{keltner1997embarrassment,fink1977humorous,clegg2012stranger,press2023humorous}, \textbf{reducing social attention} \cite{clegg2012stranger,sabini2000embarrassed,gilovich2000spotlight}, and \textbf{using indirectness} \cite{pohl2019charting,schiavo2014overt,crespo2005euphemistic,hadipragmatic}. 
Based on these strategies and XR's potential to provide novel feedback forms, we proposed a series of metaphor- and agent-based 3D feedback designs across three conversational scenarios in group conversations (someone is dominating, someone is under-participating, someone is in negative emotion).
To understand how such feedback designs are perceived in small-group conversations, we first conducted an online survey involving 50 participants, systematically comparing metaphor- and agent-based 3D feedback with direct text messages through the design probe method \cite{gaver2004cultural,graham2007probes}. Subsequently, we conducted follow-up interviews with 14 participants to gain a deeper understanding of their interpretations.

Our findings indicated that metaphor- and agent-based feedback was associated with lower perceived embarrassment and greater acceptance than the direct text messages, while maintaining usefulness and fairness. 
Interview results suggested that incorporating humor, reduced attention, and indirectness can reduce embarrassment in public feedback. Participants indicated that humor incorporation is effective when it is playful yet appropriate, reduced social attention can be supported by disconnecting feedback from individuals (e.g., through environmental cues or redirecting attention to an agent), and indirectness usage benefits from a balance between being subtle and remaining understandable. They also envisioned additional socially acceptable 3D feedback designs and interaction mechanisms to support inclusive small-group conversations.

The main contribution of our work lies in exploring how public feedback can be designed to be more socially acceptable while keeping its effectiveness for small-group conversations in XR: (1) We identified a set of design strategies, proposed corresponding 3D feedback probes to mitigate embarrassment caused by public exposure, and applied them to three conversational scenarios. (2) We conducted a two-part user study, involving an online survey and follow-up interviews, to understand how people perceive various feedback and strategies in small-group conversations and to explore any additional design opportunities. (3) We further discussed design implications to inform the future work of inclusive feedback systems for small-group conversations in XR environments.



\section{Related Work}

\subsection{Public versus Private Feedback in Small-Group Conversations}

In the context of group communication, providing feedback on performance for the group members has been proven as a powerful way to shape group dynamics \cite{villeval2020performance,samrose2018coco}. Feedback occupies a major portion of the total effort and serves an anchoring function in communication process \cite{scheidel1966feedback} in small-group conversations (i.e., three to ten equal-status people engaged in unstructured conversations \cite{bales1951channels}). 
Prior work has noted that the impact of feedback is influenced by its visibility \cite{villeval2020performance,ouverson2021analysis}, which can be presented either privately (i.e., shown only to the person to whom it applied) or publicly (i.e., shown to everyone in the group) \cite{ouverson2021analysis}. 

\textbf{Public feedback} supports social comparison \cite{festinger1954theory} and social influence \cite{argyle1957social}, which is effective for addressing problematic behavior \cite{dijkstra2008social} and has a positive effect on facilitating performance \cite{edmondson1999psychological,ancona1992bridging}. 
Despite the benefits, public corrective feedback often elicits embarrassment \cite{keltner1997embarrassment,archibald1971self}. It can single individuals out in front of others, therefore heightening concerns about negative evaluation and loss of face.
\textbf{Private feedback} is less likely to generate the social discomfort \cite{houtti2025observe} by not pointing individual out to the entire group \cite{morris2006mediating}. However, prior literature also noted that private feedback has less social influence \cite{argyle1957social} and is less effective in facilitating performance because of the absence of observers \cite{ohtsuki2015reputation}. 
Overall, these prior work highlights the trade-off between public and private feedback: while public feedback can leverage social influence to facilitate performance, it may also expose individuals to embarrassment about negative evaluation; private feedback can reduce such social discomfort, but may be less effective due to its limited visibility. 

\subsection{Existing Feedback Systems in HCI}

Prior research has extensively explored how to provide feedback on group dynamics across various settings, including video conferencing, social VR, and in-person communication \cite{bagmar2022analyzing,schmid2022does,samrose2018coco,sarda2013real,schiavo2014overt}.
Existing work has made efforts to provide feedback on a range of conversational features, including participation \cite{samrose2018coco,armstrong2024sealmates,bagmar2022analyzing,li2023improving}, turn-taking \cite{dong2012modeling,skantze2021turn}, interruption \cite{sarda2013real}, agreement \cite{janssen2007visualization,leshed2009visualizing}, and emotional valence \cite{ez2020emodash,schneider2024no}.
Meanwhile, researchers also investigated how different feedback formats shape users’ experience. For example, Schiavo et al. \cite{schiavo2014overt} conducted a study comparing overt, text-based directives with subtle directives delivered through graphical displays, and examined their effects on group dynamics. 
Beyond text and visual graphics, another playful feedback format is introducing agent as an external mediator \cite{armstrong2024sealmates,liu2021significant}. Liu et al. \cite{liu2021significant} has demonstrated that a biosignal-driven agent could support remote communication between romantic distributed partners. Similarly, Armstrong et al. \cite{armstrong2024sealmates} introduced a behavior-drive agent to promote balanced participation in video conferencing.

As an emerging medium for collaboration, XR enables group members to interact in both remote and co-located settings \cite{nguyen2020user,freeman2022working,wu2024emerging,wei2025remotechess}. Researchers have explored several approaches to provide feedback on conversational dynamics in XR. For instance, Lee et al. \cite{lee2024may} presented a multimodal attention-guidance method to facilitate turn-taking in VR-based group conversations. Yang et al. \cite{yang2023understanding} designed a virtual moderator to promote group members' balanced participation. Despite technological interventions, the social discomfort associated with public feedback may also persist in XR-based group conversations. Slater et al. compared small-group behavior in the virtual and real environment, and found that socially conditioned responses such as embarrassment could also be generated in the virtual meeting \cite{slater2000small}.

Although existing studies have explored how to promote inclusive conversations through technological assistance, most of them deliver feedback in a public way while paying little attention to the negative effects of public exposure, such as embarrassment, social pressure, and resistance \cite{edmondson1999psychological,archibald1971self,keltner1997embarrassment,gilovich2000spotlight}. Acknowledging the trade-off between public and private feedback, our work delves deeper into how public feedback in XR-based small-group conversations can be designed to be more socially acceptable while keeping its effectiveness.

\subsubsection{XR to Support Novel Feedback Forms}

XR provides more possibilities for the use of digital representations (e.g., embodied agents) and contextual cues \cite{liu2023boosting,wei2024augmented}, which demonstrates its opportunity for providing novel feedback forms in group conversations.
\textbf{One promising direction is the use of embodied agents as feedback carriers.} Prior work has identified multiple types of agent representations, including authentic, non-anthropomorphic, and other novel forms \cite{nguyen2020user}. The appearance and behavior of these virtual characters can shape users' perceptions and behaviors in XR environments \cite{nguyen2020user}. Animal-like agents, in particular, present as an emerging application in both fully and partly immersive systems. For example, Norouzi et al. \cite{norouzi2019walking} described a virtual dog in AR and found that its simulated behaviors affected participants' perception, locomotion, and social interaction with others. Similarly, Na et al. \cite{na2022mixed} leveraged a virtual cat in MR to reduce people’s mental stress and induce positive emotions.

\textbf{Another relevant direction is the use of embodied metaphors in XR.} Metaphors express ideas in a familiar and acceptable substitution for a wide range of people \cite{fernandez2006language,crespo2005euphemistic,fernandez2008sex,lubart1997emotion}. Building a metaphor helps users understand abstract concepts by relating them to something familiar or tangible \cite{carroll2003hci,jung2017metaphors}. Recent XR studies have begun to leverage this potential. For example, Wang et al. \cite{wang2018break} showed that the metaphor of ``breaking the walls'' in VR could support individual creative performance. Other work has used metaphorical interaction to help users cope with negative emotions in XR environments \cite{rasch2024mind,grieger2021trash}. Rasch et al. \cite{rasch2024mind}, for instance, enabled users to engage with representations of their thoughts by gradually organizing an inner space through the metaphor of cleaning the room''.


\subsection{Design Strategies: Reduce Embarrassment in Social Contexts}
\label{sec:strategies}

To inform more socially acceptable feedback design, several approaches to reduce embarrassment in social contexts were identified as our feedback design strategies.
\textbf{Strategy 1: Incorporate humor.} Researchers have documented humor as an effective strategy for alleviating embarrassment in interpersonal communication \cite{keltner1997embarrassment,fink1977humorous,clegg2012stranger}. 
Humor is a common remedial strategy in embarrassing interactions involving equal status individuals \cite{fink1977humorous}. 
Press and Erel \cite{press2023humorous} showed that even peripheral humorous interventions (like robotic gestures) can significantly reduce awkwardness between strangers. 
Playfulness is considered the basis of humor \cite{ruch2019humor}.

\textbf{Strategy 2: Reduce social attention.} Empirical evidence suggests that social attention can play a causal role in amplifying awkwardness: Clegg \cite{clegg2012stranger} found that inducing explicit social attention would increase reported feelings of social awkwardness. 
This aligns with the spotlight effect, in which people overestimate how much others notice their actions and appearance, particularly in embarrassing situations \cite{gilovich2000spotlight}. As a potential remedy, the appearance of a new object may capture and direct people's social attention \cite{yantis1996attentional,yantis1993stimulus}, thereby reducing their perceived social exposure.

\textbf{Strategy 3: Use indirectness.}
Prior HCI work shows that subtle interaction is typically described as socially acceptable with less embarrassment \cite{pohl2019charting}. Schiavo et al. \cite{schiavo2014overt} also found that subtle directives can influence group behavior and are more positively accepted, compared to overt, text-based directives. This rationale is consistent with linguistic and pragmatics research: indirect expressions, especially euphemisms, contribute to saving the public self-image of people in communication \cite{crespo2005euphemistic,hadipragmatic}. Metaphor is the most central device in the formation of euphemism \cite{crespo2005euphemistic,fernandez2006language}, ``in which one thing is referred to by a word or expression normally denoting another thing, idea, or action, so as to suggest some common quality shared by the two''. In light of these prior work, we formulated our Strategy 3.


\section{Designing 3D Feedback as Probes}
\label{sec:design}

Guided by the design strategies and emerging feedback forms in XR, we adopted the design probe approach \cite{gaver2004cultural,graham2007probes} and proposed a series of metaphor- and agent-based feedback probes for three conversational scenarios, aiming to mitigate the social discomfort caused by public exposure (Figure \ref{sec:design}). Drawing inspiration from prior research on human and virtual cat interaction in XR \cite{na2022mixed}, we selected a cat as the embodied agent for our agent-based feedback designs. The textual messages were also included as a baseline, reflecting the typical and direct feedback commonly used in existing systems \cite{schiavo2014overt}. Section \ref{sec:scenario} explains the rationale for selecting the conversational scenarios, while Section \ref{sec:rationale} describes how the feedback probes were developed, which we characterize by \textbf{\textit{Text (Baseline)}}, \textbf{\textit{Metaphor,}} and \textbf{\textit{Cat}}.

\subsection{Conversational Scenarios Selection}
\label{sec:scenario}

Both \textbf{balanced participation} and \textbf{positive emotions} have been identified as important and recurring themes in prior work on group conversations \cite{li2023improving,houtti2025observe,samrose2021meetingcoach,sarda2013real}.
Dominance and under-participating behavior of one or a few members are common in collaboration \cite{palmer1989controlling,latane1979many,piezon2008perceptions} and can negatively affect group interaction by limiting perspectives diversity and reducing opportunities for inclusive collaboration \cite{sargis2002informational}. 
Beyond participation balance, prior work has also examined the role of emotional valence in group conversations. Negative emotions, such as sadness, anger, and anxiety, were viewed as potentially having a negative impact on decision-making or behavioral control in group conversations \cite{murungi2019control,ferrer2016effect}. 
Drawing on this line of work, we included three representative conversational scenarios in this study: \textbf{(1) someone is dominating the conversation}, \textbf{(2) someone is under-participating in the conversation}, and \textbf{(3) someone is in negative emotion during the conversation}. The first two scenarios are associated with imbalanced participation, while the third scenario concerns emotional valence.

\subsection{Design Rationale}
\label{sec:rationale}

\begin{figure*}
    \centering
    \includegraphics[width=0.87\linewidth]{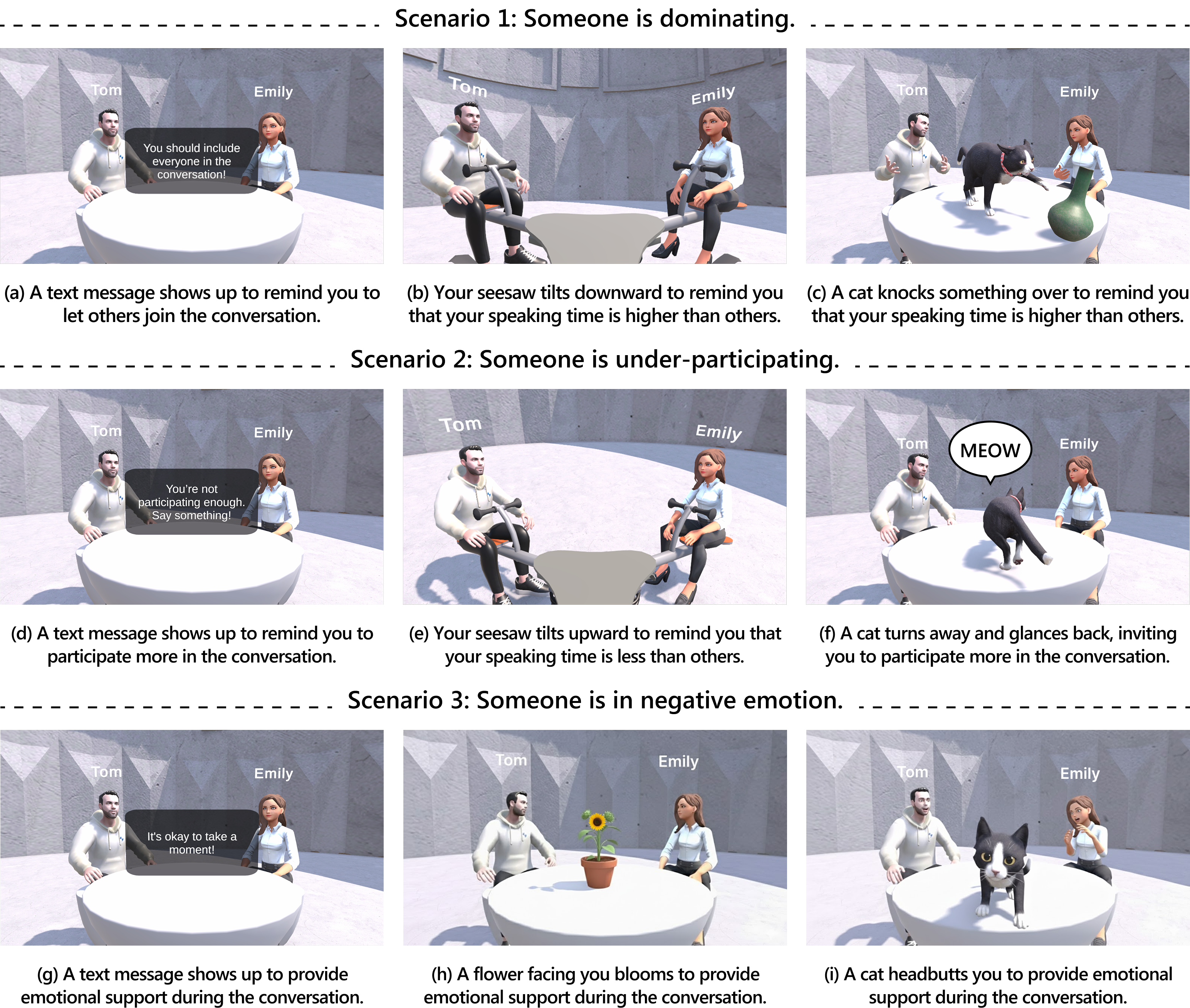}
    \caption{Overview of the feedback probes. We proposed \textbf{\textit{Text (Baseline)}}, \textbf{\textit{Metaphor}}, and \textbf{\textit{Cat}} for three conversational scenarios: someone is dominating the conversation, someone is under-participating, and someone is experiencing negative emotion. Subfigures (a), (d), and (g) show the \textbf{\textit{Text}} Baseline; (b), (e), and (h) show the \textbf{\textit{Metaphor}} probes; and (c), (f), and (i) show the \textbf{\textit{Cat}} probes.}
    \Description{Nine panels show feedback cues in a virtual conversation with two avatars seated around a table. In (a), a floating text panel reads, ``You should include everyone in the conversation!'' In (b), a seesaw tilts downward on the feedback recipient's side. In (c), a cat on the table knocks over a vase. In (d), a text panel reads, ``You're not participating enough. Say something!'' In (e), the seesaw tilts upward on the recipient's side. In (f), a cat turns away and glances back toward the recipient, with a speech bubble reading ``MEOW.'' In (g), a text panel reads, ``It's okay to take a moment!'' In (h), a potted sunflower faces the recipient and blooms. In (i), a cat leans toward the recipient for a headbutt.}
    \label{fig:design}
\end{figure*}

\begin{figure*}
    \centering
    \includegraphics[width=0.95\linewidth]{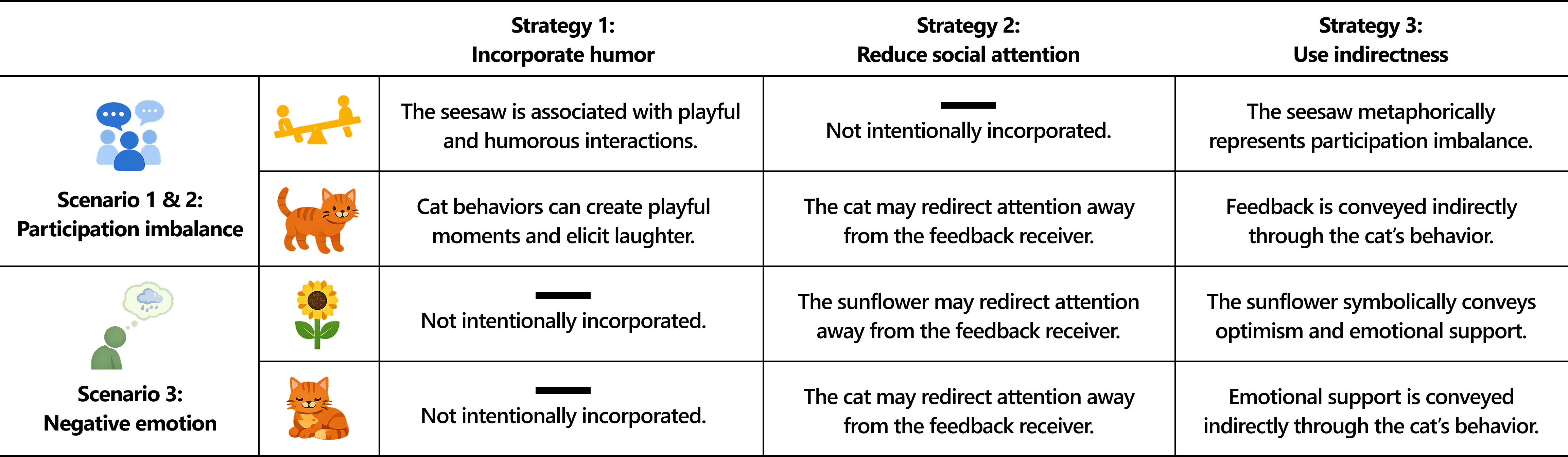}
    \caption{Overview of how the feedback probes reflect the three design strategies across the conversational scenarios. Not every probe incorporates all three strategies. An em dash indicates that a strategy was not intentionally included in the corresponding probe.}
    \Description{A matrix links feedback probes to humor, reduced social attention, and indirectness. For participation imbalance in Scenarios 1 and 2, the seesaw evokes playful interactions and metaphorically represents imbalance; reducing social attention is not intentionally incorporated. The cat incorporates all three strategies: its behavior can create playful moments and elicit laughter, may redirect attention away from the feedback recipient, and conveys feedback indirectly. For negative emotion in Scenario 3, the sunflower may redirect attention away from the recipient and symbolically conveys optimism and emotional support. The cat may also redirect attention away from the recipient and conveys emotional support indirectly through its behavior. Humor is not intentionally incorporated into either negative-emotion probe.}
    \label{fig:rationale}
\end{figure*}

\subsubsection{Text (Baseline)}

As for the Baseline, we chose explicit and straightforward textual messages, which are common found in existing systems \cite{schiavo2014overt,houtti2025observe,do2022should,kim2020bot}. We created scenario-specific text messages presented on a 3D UI panel (\textit{``You should include everyone in the conversation!''} for Scenario 1 \cite{schiavo2014overt}, \textit{``You are not participating enough. Say something!''} for Scenario 2 \cite{schiavo2014overt}, \textit{``It's okay to take a moment.''} for Scenario 3), to compare with the following designs. 

\subsubsection{Metaphor}

\textbf{Scenario 1 and 2: Participation imbalance.}
Previous research has used the seesaw to represent imbalance \cite{patel2013visual,mrstik2026you} (Strategy 3: Use indirectness). In a study designing visual metaphors for providing feedback to clinicians, Patel et al. \cite{patel2013visual} found that seesaw tilting offered an intuitive and straightforward way to identify who was dominating the conversation.
Additionally, introducing a novel object can redirect people's attention away from the individual \cite{yantis1996attentional,yantis1993stimulus} (Strategy 2: Reduce social attention).
Moreover, the seesaw is associated with humor and playful interactions, particularly during moments of togetherness among peers \cite{gunnarsdottir2025moments} (Strategy 1: Incorporate humor).
Here seesaw tilting was used to indicate participation imbalance in Scenario 1 and 2. Specifically, when someone is dominating the conversation, the increased participation was mapped onto increased weight on the seesaw. When someone is under-participating, this lower level of participation was represented by having that person’s side rise upward.

\textbf{Scenario 3: Negative emotion.}
For Scenario 3, the sunflower was chosen as a metaphorical representation for providing emotional support. In everyday language, sunflowers are seen as a symbol of optimism, happiness, and good luck \cite{scott2026flower,lestari2023decoding} (Strategy 3: Use indirectness). Beyond its metaphorical meaning, the sunflower also functions as an attentional cue: as a novel visual element, it may draw social attention away from the individual experiencing negative emotion \cite{yantis1996attentional,yantis1993stimulus} (Strategy 2: Reduce social attention).

\subsubsection{Cat}
Research on companion animals shows that pets' behaviors can provide a greater source of laughter for people \cite{valeri2006tails} (Strategy 1: Incorporate humor).
In Scenario 1, the cat knocks over nearby objects to attract the dominant participant’s attention and subtly remind them to make space for others. In Scenario 2, the cat turns away and then looks back toward the under-participating participant, using this gesture as a gentle invitation to engage in the conversation. 
In Scenario 3, when someone is in negative emotion during the conversation, a virtual cat will start headbutting and seek for petting.
In all the scenarios, the feedback is conveyed indirectly through the cat’s behavior rather than through an explicit message (Strategy 3: Use indirectness). By directing attention toward the cat, these behaviors may also reduce the extent to which the feedback recipient becomes the focus of the group’s attention, helping them feel less singled out (Strategy 2: Reduce social attention).

\section{Survey}
\label{sec:survey}

To gather participants' initial perceptions of these feedback probes, we conducted an online survey. The survey was distributed through Reddit communities related to XR and group communication, to reach a broader pool of population. The first 30 participants who completed the survey received \$15 in compensation, while the remaining participants were entered into a raffle for a chance to receive \$50.

\subsection{Survey Participants}
A total of 50 participants completed the survey. Participants ranged in age from 19 to 44 years (M = 29.0, SD = 6.3). 29 participants identified as male (58\%) and 21 identified as female (42\%). In terms of race and/or ethnicity, participants identified as Black or African American (42\%), White (32\%), Asian (20\%), Hispanic or Latino (4\%), and other racial or ethnic backgrounds (2\%).

\subsection{Survey procedure}

The survey was hosted on the Qualtrics online platform and took approximately 30 minutes to complete. It is composed of screening, information sheet, demographic information and the main survey. First, the participants answered three screening questions about age, ability to read from a computer screen with normal vision, and ability to understand basic written English. Eligible participants were able to view and download an information sheet that provides a detailed description of the survey study (a waiver of documented consent was approved by the Institutional Review Board). Participants who agreed to participate then proceeded to the remaining sections.

\textbf{Demographic information} gathered information about participants’ age, gender, race and/or ethnicity, prior experiences in small-group conversations, and prior usage of XR technologies.

\begin{figure*}
    \centering
    \includegraphics[width=0.98\linewidth]{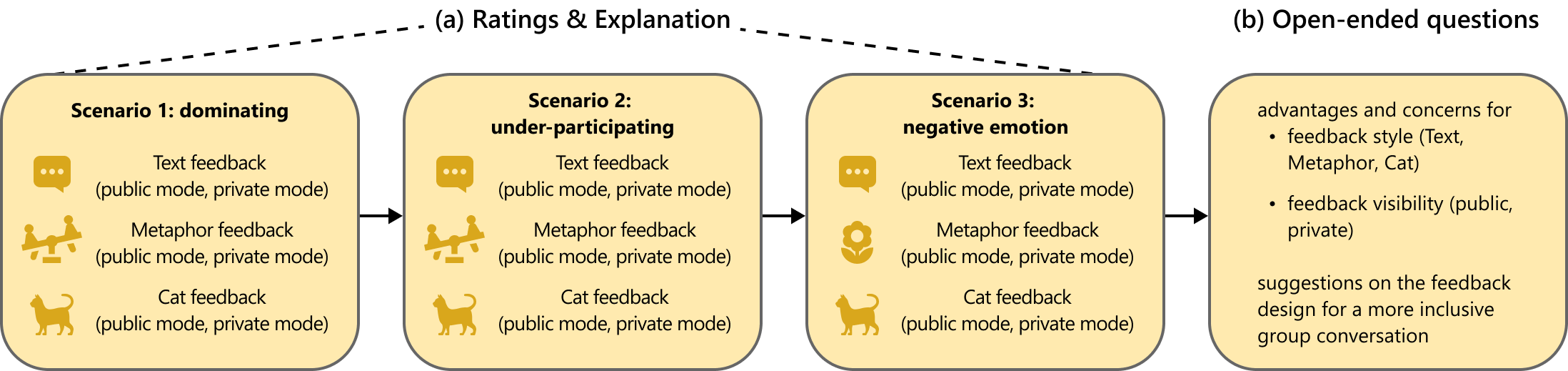}
    \caption{Flowchart of the main survey procedure. (a) Participants were presented with three small-group conversation scenarios. For each scenario, they provided ratings and written explanation for three feedback styles—Text (Baseline), Metaphor, and Cat—with each style shown in both public and private modes. (b) Participants answered open-ended questions about the perceived advantages and concerns of different feedback styles and visibility, and offered suggestions for designing socially acceptable feedback to support more inclusive group conversations.}
    \Description{A flowchart contains four rounded boxes connected by arrows from left to right. The first three boxes are grouped under ``Ratings and Explanation'' and represent domination, under-participation, and negative emotion. Each scenario box lists Text, Metaphor, and Cat feedback, each with public and private modes. Speech bubbles represent Text feedback, and cat icons represent Cat feedback. The metaphor icon is a seesaw for the first two scenarios and a sunflower for the third. The fourth box, labeled ``Open-ended questions,'' contains prompts about advantages and concerns regarding feedback styles and visibility, followed by suggestions for feedback design.}
    \label{fig:surveyprocedure}
\end{figure*}

\begin{figure*}
    \centering
    \includegraphics[width=1.0\linewidth]{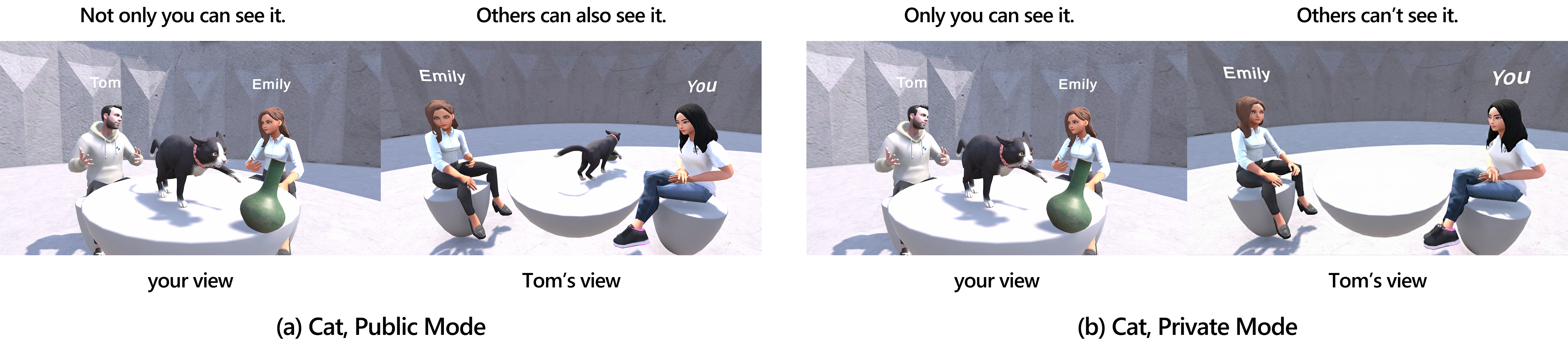}
    \caption{Illustration of feedback visibility in the survey (take Cat feedback as an example). We presented side-by-side views in the survey, displaying both the participant’s own perspective (your view) and another group member’s perspective (Tom’s view). (a) In the public mode, the cat is visible in both your view and Tom's view. (b) In the private mode, the cat is only visible in your view and does not appear in Tom's view.}
    \Description{Four screenshots compare public and private Cat feedback from two viewpoints in a virtual group conversation. Each pair shows the recipient's view on the left and Tom's view on the right. The recipient sees Tom and Emily across a round table, while Tom sees Emily and the recipient. In the public-mode pair (a), a cat appears on the tabletop in both views. In the private-mode pair (b), the cat remains visible on the tabletop in the recipient's view, but the tabletop is empty in Tom's view.}
    \label{fig:visibility}
\end{figure*}

\textbf{Main survey} consisted of three scenarios and an open-ended question session (Figure~\ref{fig:surveyprocedure}). To mitigate potential order effects, the presentation order of the three scenarios was randomized for each participant. At the beginning of the survey, participants were asked to imagine that they were taking part in a small-group conversation with two other peers. For each scenario, participants first watched a short video depicting the situation and then evaluated three feedback designs (\textbf{\textit{Text}}, \textbf{\textit{Metaphor}}, and \textbf{\textit{Cat}}). Each design was shown under two visibility conditions: public and private (Figure~\ref{fig:visibility}). Specifically, in the private mode, only the participant could see the feedback, while in the public mode, all group members could see it.

After viewing each design, participants rated their perceptions using 5-point Likert-scale. We measured their perceived usefulness (\textit{``It would help reduce dominant participation/encourage quieter speakers to contribute more/ease negative emotions during the conversation.''}) \cite{samrose2021meetingcoach,do2022should,kim2020bot}, fairness (\textit{``It would help us participate more openly and fairly.''}) \cite{kim2020bot}, comfort (\textit{``I would feel comfortable receiving it during a conversation.''}) \cite{kim2020bot}, embarrassment (\textit{``Receiving this feedback would make me feel embarrassed.''}) \cite{do2023designing}, and distraction (\textit{``I would be distracted by it.''}) \cite{schiavo2014overt}. 
We additionally included questions asking participants what aspects of the design contributed to reducing embarrassment. Specifically, we included three items: Strategy 1 (\textit{``It would be playful and easier to accept.''}), Strategy 2 (\textit{``It would avoid putting me at the center of attention.''}), and Strategy 3 (\textit{``It would be indirect and help me save my face.''}).
After completing three scenarios, participants answered open-ended questions about perceived advantages, concerns, and design suggestions for the solutions. The metrics and corresponding survey questions are listed in Appendix~\ref{survey_q}.

\subsection{Data Collection and Analysis}

We conducted a validation process to remove responses that have duplicated IP addresses and incomplete results.
The quantitative survey ratings were analyzed using descriptive statistics and inferential testing. 
As our data was collected on a 5-point Likert scale and each participant evaluated all the feedback probes, we conducted Friedman tests \cite{marshall2011introduction} to compare \textbf{\textit{Text}}, \textbf{\textit{Metaphor}}, and \textbf{\textit{Cat}} for each dependent variable (usefulness, fairness, comfort, embarrassment, and distraction).
For the qualitative open-ended responses, we conducted thematic analysis  \cite{braun2006using}: two researchers independently coded participants' responses, and three researchers then grouped similar codes into broader themes.

\section{Survey Results}
\label{sec:surveyresults}

\begin{figure*}
    \centering
    \includegraphics[width=0.98\linewidth]{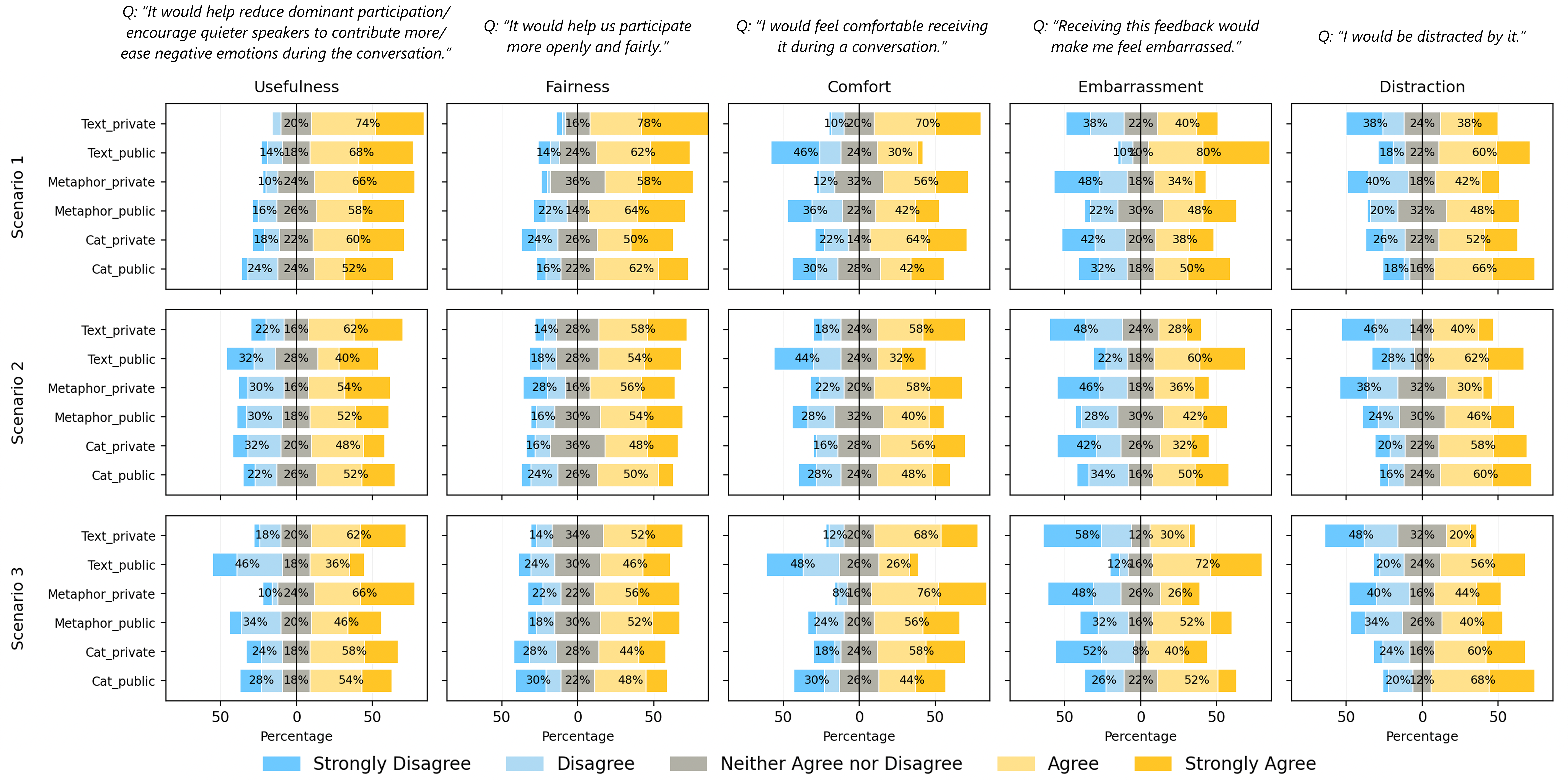}
    \caption{Under private and public conditions, participants' ratings of \textbf{\textit{Text}}, \textbf{\textit{Metaphor}}, and \textbf{\textit{Cat}} across three scenarios on usefulness, fairness, comfort, embarrassment, and distraction. Bars show the percentage distribution of responses on a 5-point Likert scale.}
    \Description{Fifteen diverging stacked bar charts are arranged in three scenario rows and five rating columns. Rows represent domination, under-participation, and negative emotion. Columns show usefulness, fairness, comfort, embarrassment, and distraction. Each panel contains six bars, ordered as Text private, Text public, Metaphor private, Metaphor public, Cat private, and Cat public. Disagreement extends to the left, agreement to the right, and neutral responses are centered. Segments represent the five response categories from strongly disagree to strongly agree, with percentage labels summarizing responses. Usefulness and fairness ratings generally lean toward agreement. Across all three scenarios, private feedback receives more agreement for comfort and less agreement for embarrassment than the corresponding public feedback. Within public delivery, Metaphor and Cat receive more agreement for comfort and less agreement for embarrassment than Text. Agreement that public Text feedback would cause embarrassment is 80 percent, 60 percent, and 72 percent for Scenarios 1, 2, and 3, respectively. Cat receives more agreement for distraction than Metaphor in every scenario and visibility condition.}
    \label{fig:results}
\end{figure*}


\subsection{Public vs. Private Feedback Delivery}

Across all the feedback designs and scenarios, participants consistently preferred private over public delivery on socially sensitive dimensions. As shown in Figure~\ref{fig:results}, public delivery elicited higher \textbf{embarrassment} and lower \textbf{comfort} than private delivery for each design. This directly supports our motivation: publicly delivered feedback is more likely to introduce social discomfort in small-group conversations. A similar contrast emerged for \textbf{distraction}. Across all designs, public delivery was perceived as more distracting than private delivery, which suggests that making feedback visible to the group may interrupt the conversation flow.

\subsection{Preference for Text, Metaphor, and Cat}

Figure~\ref{fig:results} summarizes participants’ ratings of the \textbf{\textit{Text}} (baseline), \textbf{\textit{Metaphor}}, and \textbf{\textit{Cat}} probes across three scenarios, reporting perceived usefulness, fairness, comfort, embarrassment, and distraction under both private and public conditions. Overall, these results highlight that metaphor- and agent-based designs can reduce some of the public-delivery costs.

\subsubsection{Usefulness and Fairness}
In Scenario 1 and 2, \textit{Text\_private} received the highest positive ratings, with average percentage of positive responses (agree and strongly agree) of Scenario 1 at 74\% and Scenario 2 at 62\%. Whereas in Scenario 3, \textit{Metaphor\_private} was rated the most useful (66\%).
Fairness ratings were generally positive or neutral across \textbf{\textit{Text}}, \textbf{\textit{Metaphor}}, and \textbf{\textit{Cat}} under private and public conditions, with no clear differentiation across scenarios.

\subsubsection{Comfort and Embarrassment}
\textit{Text\_public} consistently received the highest embarrassment and lowest comfort ratings in every scenario (comfort: 26--32\%; embarrassment: 60--80\%). To further examine differences among the three public probes, we conducted Friedman tests for each scenario. For comfort, the Friedman test showed a significant difference among \textbf{\textit{Text}}, \textbf{\textit{Metaphor}}, and \textbf{\textit{Cat}} ($\chi^2 = 6.561$, $p < .05$) in Scenario 1. The Wilcoxon signed-rank tests with Bonferroni correction showed that \textbf{\textit{Text}} was rated as significantly less comfortable than \textbf{\textit{Metaphor}} ($p < .05$) and \textbf{\textit{Cat}} ($p < .05$). The difference between \textbf{\textit{Metaphor}} and \textbf{\textit{Cat}} was not significant ($p = 1.000$). A significant difference in comfort was also found in Scenario 3 ($\chi^2 = 18.536$, $p < .001$). Post-hoc comparisons showed that \textbf{\textit{Text}} was rated as significantly less comfortable than \textbf{\textit{Metaphor}} ($p < .001$), whereas the differences between \textbf{\textit{Text}} and \textbf{\textit{Cat}} ($p = .113$), and between \textbf{\textit{Metaphor}} and \textbf{\textit{Cat}} ($p = .184$), were not significant.
For embarrassment, the Friedman test showed a significant difference among the three public probes in Scenario 1 ($\chi^2 = 14.683$, $p = .001$). Post-hoc analysis showed that \textbf{\textit{Text}} was rated as significantly more embarrassing than \textbf{\textit{Metaphor}} ($p = .001$) and \textbf{\textit{Cat}} ($p < .005$). The difference between \textbf{\textit{Metaphor}} and \textbf{\textit{Cat}} was not significant ($p = 1.000$).

\subsubsection{Distraction}
Overall, public delivery was perceived as more distracting than private delivery. Under public delivery, \textbf{\textit{Text}} and \textbf{\textit{Cat}} were rated as the most disruptive (\textit{Text\_public}: 60\% in Scenario 1, 62\% in Scenario 2, 56\% in Scenario 3; \textit{Cat\_public}: 66\% in Scenario 1, 60\% in Scenario 2, 68\% in Scenario 3)., while \textbf{\textit{Metaphor}} was consistently rated as the least distracting (\textit{Metaphor\_public}: 48\% in Scenario 1, 46\% in Scenario 2, 40\% in Scenario 3). The Friedman test showed a significant difference among the three probes in Scenario 3 ($\chi^2 = 10.161$, $p < 0.05$). Post-hoc analysis showed that \textbf{\textit{Metaphor}} was rated as significantly less distracting than \textbf{\textit{Cat}} ($p < 0.05$). There was no significance between \textbf{\textit{Text}} and \textbf{\textit{Metaphor}} ($p = 0.091$), as well as \textbf{\textit{Text}} and \textbf{\textit{Cat}} ($p = 1.000$) in this scenario.

\subsubsection{In-depth Feedback}

In open-ended survey responses, participants described perceived advantages, concerns, and suggested improvements associated with the feedback probes (see Appendix~\ref{survey_op} for specific themes and participant counts). \textbf{\textit{Text}} (baseline) was valued for clarity and directness, yet perceived as more likely to cause embarrassment. \textbf{\textit{Metaphor}} was seen as less intrusive but sometimes ambiguous or insufficiently informative. \textbf{\textit{Cat}} was described as funny, attractive, emotionally low-pressure, and as helpful for easing awkwardness in public conditions, but limited by lower information density for serious contexts. Participants also suggested offering flexible options that allow users to choose their preferred probe incorporating multimodal cues such as haptic pulses and voice reminders.

\section{Interview}
\label{sec:interview}

To gain a deeper understanding of participants’ survey responses and elicit additional suggestions, the research team invited a subset of survey participants to take part in follow-up interviews. 

\subsection{Interview Participants}

\begin{table}[t]
  \centering
  \caption{Demographic information of the participants (N=14).
  Each participant's background information was summarized
  based on their responses in the survey.}
  \label{tab:demographic}
  \small
  \resizebox{\linewidth}{!}{%
    \begin{tabular}{@{}cccccc@{}}
      \toprule
      \textbf{ID}
      & \textbf{Age}
      & \textbf{Gender}
      & \shortstack{\textbf{Race and/or}\\\textbf{Ethnicity}}
      & \shortstack{\textbf{Frequency of}\\
                    \textbf{Small-group Conversations}}
      & \shortstack{\textbf{Prior XR}\\\textbf{Experience}}\\
      \midrule
      P1  & 24 & Male   & Asian
          & At least three times a week & yes\\
      P2  & 20 & Female & Asian
          & At least one time a week & yes\\
      P3  & 44 & Male   & White
          & Almost every day & yes\\
      P4  & 24 & Male   & Asian
          & At least three times a week & no\\
      P5  & 19 & Female & White
          & Less than one time a week & no\\
      P6  & 27 & Female & Asian
          & At least three times a week & yes\\
      P7  & 27 & Male   & Black or African American
          & At least once a week & yes\\
      P8  & 23 & Male   & Black or African American
          & At least once a week & yes\\
      P9  & 19 & Male   & Asian
          & At least three times a week & yes\\
      P10 & 22 & Female & Asian
          & At least three times a week & yes\\
      P11 & 23 & Female & Asian
          & At least three times a week & yes\\
      P12 & 27 & Female & Asian
          & At least one time a week & no\\
      P13 & 19 & Male   & Hispanic or Latino
          & At least one time a week & yes\\
      P14 & 27 & Male   & Asian
          & At least one time a week & yes\\
      \bottomrule
    \end{tabular}%
  }
\end{table}

We recruited follow-up interview participants from those who had completed the survey. Participants were invited if they met the following criteria: (1) they selected “yes” to the question asking whether they were interested in a future interview, and (2) they provided relatively detailed and thoughtful responses to the survey’s open-ended questions.

A total of 14 participants took part in the follow-up interviews (gender: 6 female, 8 male; age: median = 23.5, mean = 24.64, SD = 6.37). As shown in Table~\ref{tab:demographic}, 9 participants identified as Asian, 2 as Black or African American, 2 as White, and 1 as Hispanic or Latino. Most participants reported engaging in small-group conversations (e.g., meetings, discussions, and social gatherings) at least once a week, with only one participant reporting participation less than once a week. Regarding prior XR experience, 11 participants reported having prior experience, whereas 3 reported no prior experience. Each participant was compensated \$15 for their time.

\subsection{Interview Procedure}

The follow-up interviews were conducted remotely over Zoom, with each session lasting approximately 40-60 minutes. Each session involved one participant and two researchers: one researcher leading the interview, while the other researcher took notes and provided assistance as the participant needed. Before the interview began, we obtained participants’ consent for audio recording and reminded them that they could revise or add nuance to their survey responses during the interview. We also informed participants that their opinions during the interview did not need to be consistent with the ratings they had given in the survey, and they were free to share any additional, updated, or more nuanced thoughts. The interview consisted of three parts, as shown in appendix~\ref{interview_q}.
The interview consisted of three parts, including \textbf{1) revisiting public versus private feedback in small-group conversations}, \textbf{2) discussing \textit{Text}, \textit{Metaphor}, and \textit{Cat} feedback probes}, and \textbf{3) eliciting additional design suggestions}. First, participants discussed their views on public versus private feedback based on their earlier survey responses. Second, participants explained their perceptions of these feedback designs across several dimensions, including usefulness, fairness, distraction, comfort, and embarrassment. We also asked how they interpreted the design strategies underlying these probes. Third, participants provided additional design suggestions for the socially acceptable 3D feedback design. The full interview protocol is provided in Appendix~\ref{interview_q}.

\subsection{Data Collection and Analysis}

All audio recordings were transcribed into text paragraphs using Zoom’s automatic transcription function and then manually corrected. 
In analyzing the data, we adopted the thematic analysis approach. Two researchers first independently reviewed the transcripts and assigned initial codes to meaningful segments of the data. They then compared coding results and iteratively grouping them into higher-level categories and themes. Three researchers further discussed the coding outcomes, refined the thematic structure, and reached consensus on the final themes.

\section{Interview Results}
\label{sec:interviewresults}

In general, the interview results extended the survey findings by revealing more nuanced and contextual considerations and clarifying how the feedback probes were perceived to reduce embarrassment and could be further refined.

\subsection{Contextual Preferences for Feedback Visibility}
\label{sec:7.1}

Participants emphasized that preferences for public versus private feedback depend on contextual factors, including feedback valence, group chemistry, and behavioral context.
\textbf{Feedback valence.}
Participants noted that positive feedback is more socially acceptable and encouraging when shared publicly (P1, P3, P5), such as visible reactions like \textit{``thumb-up''} or \textit{``claps''} (P1). Similarly, P3 stated that \textit{``positive reinforcement would work well in public feedback''} because \textit{``everybody knows you're being encouraged''}, while more critical feedback should remain private to avoid discomfort.
\textbf{Group chemistry.}
Participants mentioned that public feedback feels more acceptable in groups with higher level of familiarity and trust (P3, P5). For example, P3 noted that receiving feedback in public is easier \textit{``with friends that you trust''}, whereas \textit{``it can scare the individual and hinder improvement in unfamiliar settings''}. 
\textbf{Behavioral context.}
Participants highlighted that feedback visibility should adapt to the context of behavior. Public feedback was preferred when behaviors disrupted group progress or required collective awareness (e.g., off-topic discussion), whereas private feedback was seen as more appropriate for sensitive situations such as low participation due to stress or lack of confidence (P1). Participants also noted that feedback visibility could be influenced by individual preferences, social roles and power dynamics (P11), and formal or informal settings (P10).

\subsection{Perceptions of Text, Metaphor, and Cat}
\label{sec:7.2}

\subsubsection{Usefulness and Fairness}
Under public delivery condition, participants generally perceived \textbf{\textit{Text}}, \textbf{\textit{Metaphor}}, and \textbf{\textit{Cat}} as effective in helping users adjust their behavior and supporting fair and open conversation.
\textbf{\textit{Text (Baseline)}} was viewed as a direct form of feedback, allowing users to \textit{``understand what to do next immediately''} (P1, P5, P6).
\textbf{Metaphor-based designs} were also considered useful for addressing participation imbalances (P2, P10) and providing emotional support (P4, P5, P8, P10). The seesaw was regarded as effective reminder because it offered a \textit{``perspective change''} (P2), while the sunflower blooming \textit{``creates a pleasant atmosphere''} (P4), \textit{``brings the sense that something good is happening''} (P8), and \textit{``is a soft, implicit, but powerful universal cultural symbol''} (P3).
\textbf{Cat-based designs} were described as capable of \textit{``breaking the dominant person's flow of conversation''} (P2), \textit{``encouraging people to participate in the conversation''} (P3, P5), and easing negative emotions because of pets' emotional support characteristics. (P4, P6, P9).
Despite these positive perceptions, some participants considered the metaphor- and cat-based designs indirect or somewhat vague, noting that they might take time to interpret or could confuse users (P3, P5, P14).

\subsubsection{Comfort and Embarrassment}
\textbf{\textit{Text (Baseline)}} was often perceived as socially intrusive and emotionally harsh, with participants describing it as \textit{``cold''}, \textit{``like a warning''} (P2), \textit{``a sense of blame and criticism''} (P10), and embarrassing when publicly visible (P3, P4, P5, P6, P8, P9, P11, P14). 
However, participants consistently emphasized that \textbf{metaphor- and cat-based designs} provide a more comfortable and socially acceptable experience compared to \textbf{\textit{Text}} in public delivery. This was largely attributed to their indirect and less confrontational nature (P1, P3, P4, P5, P6, P8, P9, P10, P14). As P1 said, \textit{``Cat and Metaphor let me feel more comfortable, because they are expressed in an indirect and interesting way.''} P4 similarly noted, \textit{``In the negative-emotion scenario, flowers and cat are both a soothing way, because they don’t expose your negative emotions directly to others.''}

\subsubsection{Distraction}
\textbf{\textit{Text}} can be distracting and interruptive, as it demands cognitive effort to read text (P1, P6) and can disturb the flow of the conversation (P3, P5).
\textbf{\textit{Metaphor}} is less distracting due to its more subtle nature (P3, P4, P5, P6, P10), while animations of \textbf{\textit{Cat}} can distract people from the conversations (P1, P2, P8, P9, P10, P13). Importantly, participants noted that distraction is not necessarily a negative thing. For example, distraction can be beneficial when it interrupts undesirable behavior, such as helping to \textit{``distract the person who is dominating the conversation''} (P2).
Additionally, agent-based elements such as the cat can create shared conversational entry points, with P3 describing how it \textit{``gives other people a perfect excuse to enter the conversation,''} such as commenting on the cat itself, thereby facilitating turn-taking. In negative emotion situations, distraction could be \textit{``a kind of distraction and break''} when experiencing negative emotions (P9).

\subsection{Perceptions and Extensions of Design Strategies for Reducing Embarrassment}
\label{sec:7.3}

Participants reflected on how design strategies were realized in the proposed designs and contributed to reducing embarrassment caused by public exposure. Using the proposed \textbf{\textit{Metaphor}} and \textbf{\textit{Cat}} probes as starting points, participants also envisioned additional socially acceptable 3D feedback designs and interaction mechanisms.

\subsubsection{Strategy 1: Incorporate Humor}
\label{sec:7.3.1}
Participants highlighted humor as an important strategy for reducing embarrassment by softening the tone and making it easier to accept. Both metaphor- and cat-based designs were perceived as incorporating humor through their \textit{``funny and interesting''} presentation, which avoids treating feedback as \textit{``a very serious or negative thing''} (P1). P9 expressed that \textit{``Seesaw is humorous and playful because it's an activity which we would conduct in the playground or park in our daily lives''}. Similarly, some participants expressed that cat feedback can remind people in friendly way and are easier to accept because the cat is adorable (P1, P3, P4, P6) and common in life (P2, P5). \textit{``It is cute and adorable and it will serve as a friendly way to make me notice the problem''} (P6). However, some participants also suggested a need of moderate humor in representation, as if it is purely humorous, it will have negative impact on the meeting content (P1).

\begin{figure}[h]
    \centering
    \includegraphics[width=0.8\linewidth]{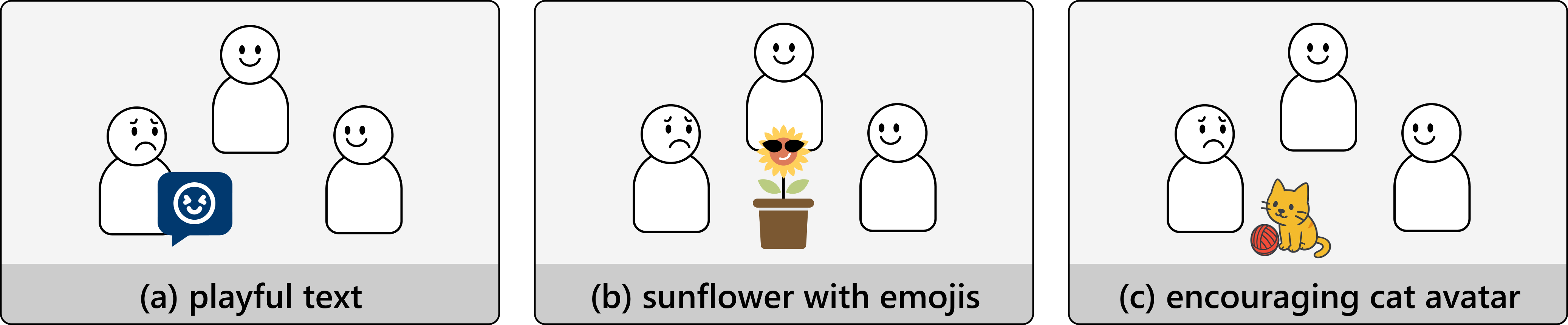}
    \caption{Participants envisioned how \textbf{Strategy 1 (Incorporate humor)} can be realized and extended through other designs, illustrated by coauthor. When someone is in negative emotion during the conversation, (a) a joke is told through text message; (b) a sunflower with glasses and emojis to provide emotional support; (c) a cat carries a small ball next to the person who has negative emotion, drops the ball with encouragement.}
    \Description{}
    \label{fig:humor}
\end{figure}

Participants envisioned that humor can be further realized through more expressive and engaging design elements (Fig. \ref{fig:humor}). 
For metaphor-based designs, participants proposed incorporating more imaginative and lighthearted elements (e.g., a sunflower with glasses as shown in Fig. \ref{fig:humor}(b)) to create a playful tone (P10). 
For agent-based feedback, they emphasized playful and interactive behaviors, such as exaggerated animations or actions (e.g., a cat carrying an encouraging object as shown in Fig. \ref{fig:humor}(c)), which make feedback feel lighter and more approachable (P2, P3, P8, P10).


\subsubsection{Strategy 2: Reduce Social Attention} 
\label{sec:7.3.2}
Participants discussed reducing attention as a way to mitigate embarrassment by shifting focus away from individuals. While some subtle representations (e.g., small metaphorical changes) were described as \textit{``small and not so distinct sign''} (P6) and \textit{``not a big issue''} (P9), participants noted that current designs of metaphor still carry a pointing effect, making it obvious who the feedback is directed at. For example, visible changes such as position change or noticeable animations can still draw group attention to a specific individual, limiting their ability to reduce social pressure (P1, P6). 

\begin{figure}[h]
    \centering
    \includegraphics[width=0.8\linewidth]{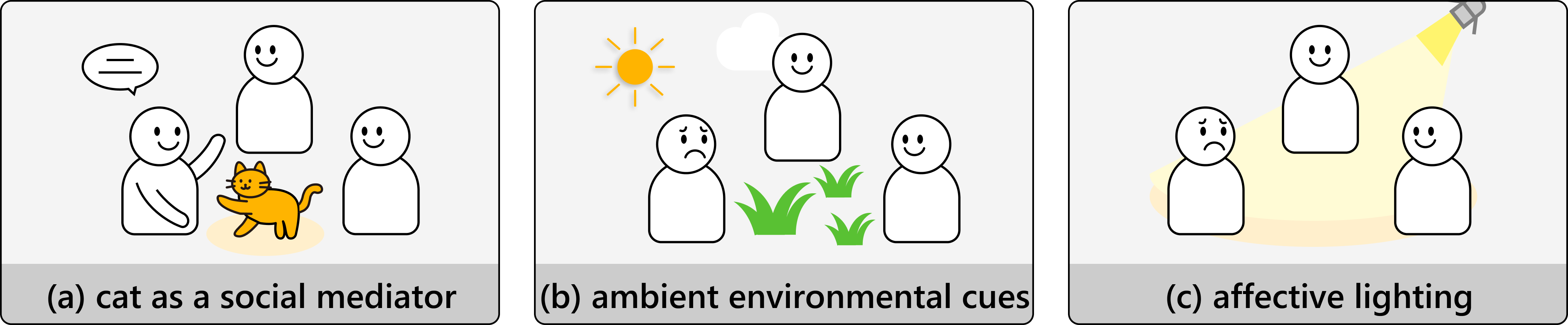}
    \caption{Participants envisioned how \textbf{Strategy 2 (Reduce social attention)} can be realized and extended through other designs, illustrated by coauthor. (a) When someone is dominating the conversation, a cat acts as a social facilitator in the group, coordinating turn-taking and guiding participation. (b) When someone is in negative emotion, use environmental cues such as sunlight and grassland to create a more positive atmosphere. (c) When someone is in negative emotion, use color to help participants feel better.}
    \Description{}
    \label{fig:attention}
\end{figure}

Some participants suggested alternative approaches that disconnect feedback from the individual, such as through environmental cues or by redirecting attention to an agent (Fig. \ref{fig:attention}). For metaphor-based designs, participants proposed feedback as environmental or ambient changes (e.g., color, lighting, weather, or scene transitions), allowing users to perceive feedback without being directly pointed out (P1, P2, P5, P8). For agent-based designs, participants suggested that the agent could delivers feedback while attracting attention to itself. For example, the agent may take on roles such as a social facilitator or an additional group member, coordinating turn-taking and guiding participation (P6). By drawing attention toward the agent rather than any individual, this approach helps reduce social attention directed at individuals.


\begin{figure}[h]
    \centering
    \includegraphics[width=1.0\linewidth]{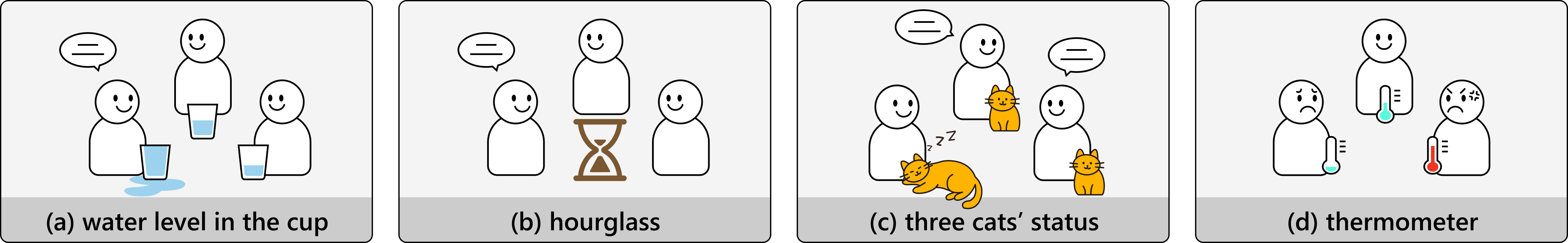}
    \caption{Other designs envisioned by participants about how \textbf{Strategy 3 (Use indirectness)} can be realized and extended. (a) Use the water level in a cup to represent each member's engagement. (b) An hourglass to signal that a dominant participant’s speaking time is running out. (c) There are three cats in the group, and each cat's status reflects the corresponding group member's participation level. (d) Use a thermometer to represent each member's emotional status.}
    \Description{}
    \label{fig:indirect}
\end{figure}

\begin{figure}[h]
    \centering
    \includegraphics[width=1.0\linewidth]{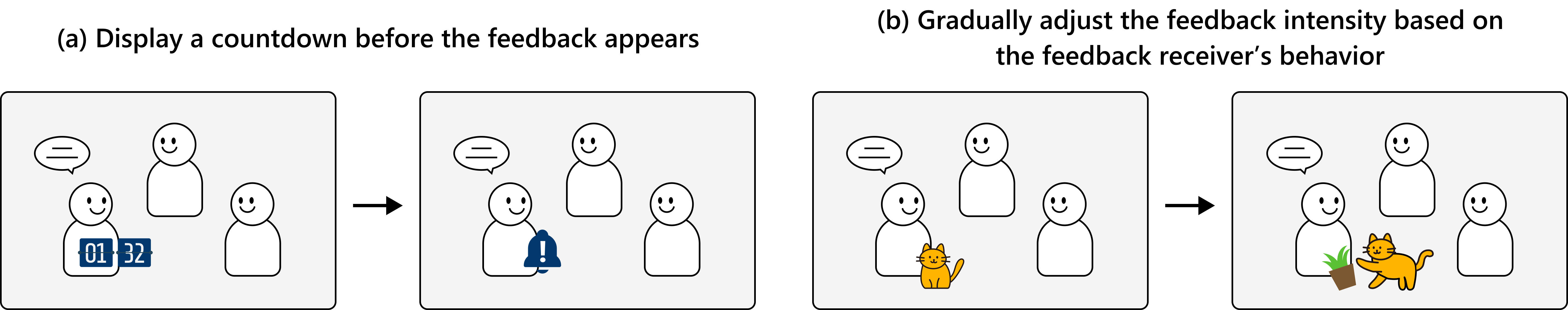}
    \caption{Participants also mentioned that there should be a transition for delivering public feedback. Here we take Scenario 1 (someone is dominating the conversation) as an example. (a) A countdown is shown before the public feedback appears, allowing the feedback receiver time to make changes or get prepared for the upcoming reminder. (b) The feedback begins soft and becomes stronger if the feedback receiver continues the dominant behavior. Specifically, the cat simply wags its tail to remind the person. If the dominant person is still talking after a while, the cat will exhibit more aggressive behaviors like knocking something over.}
    \Description{}
    \label{fig:transition}
\end{figure}

\subsubsection{Strategy 3: Use Indirectness}
Participants also discussed indirect feedback as a way to mitigate embarrassment by conveying feedback in a less explicit manner. They described metaphor and cat-based feedback as \textit{``indirect and interesting''} (P1) and \textit{``friendly''} without making users feel \textit{``offended or pressured''} (P2). However, participants also emphasized the need for a balance between indirectness and clarity (P5). As P5 suggested, \textit{``We need something not too vague, but vague enough''} so that it takes other people some time to process what's going on. If feedback is too indirect, as P1 noted, \textit{``it will confuse people''}, and others pointed out that metaphorical cues may require prior knowledge to interpret (P2, P8).

Participants envisioned that indirectness can be achieved by representing feedback through subtle forms rather than explicit indicators (Fig. \ref{fig:indirect}). For metaphor-based designs, participants proposed using visual metaphors such as water levels, writing amount, hourglasses, or microphones to guide inclusive participation (P4, P5, P10, P13). For emotional states, participants suggested affective metaphors (e.g., temperature, color, lighting, or environmental scenes) to subtly convey feelings and create a more supportive atmosphere (P1, P2, P5, P8, P14).
For agent-based designs, the agent's behavior or status may reflect participation or emotional conditions (P4, P9, P10, P14).

\subsubsection{Additional Interaction Mechanisms}
\label{sec:7.3.4}

Participants also proposed additional mechanisms for delivering the feedback. 
\textbf{Temporal Delivery and Transition.} Participants mentioned that there should be a transition for delivering public feedback (Fig. \ref{fig:transition}). P14 proposed that feedback could be delivered gradually, allowing time for self-correction or increasing feedback intensity if the behavior continues.
\textbf{Personalization.} Participants emphasized the importance of personalizing appearance to improve comfort and engagement. For instance, they suggested offering a range of alternative agents (e.g., different animals or characters) and allowing users to select their preferred option, which can make feedback feel more positive and less intrusive (P1, P3, P4, P8, P10).

\section{Discussion}

Based on the findings, we further discuss design implications for future socially acceptable feedback design in XR-based small-group conversations.

\subsection{Constructing Embodied Agents and Metaphors as XR Feedback Forms}


Previous work has explored using embodied agents and metaphors to shape users’ perceptions and behaviors in XR environments \cite{nguyen2020user,norouzi2019walking,na2022mixed,wang2018break,rasch2024mind,grieger2021trash}. Our work leveraged this potential and constructed embodied agents and metaphors as novel feedback forms in XR. 
Participants responded positively to this. They described metaphor- and cat-based feedback as useful for encouraging behavioral adjustment, less confrontational, and more supportive (Section \ref{sec:7.2}). These responses suggest that feedback forms may shape how a reminder is socially interpreted. For example, participants interpreted the cat’s behavior as an invitation to participate or a source of emotional support, giving the feedback a more approachable tone.
Participants also envisioned a broader range of metaphorical representations, agent behaviors, and environmental cues (Section \ref{sec:7.3}). 
Overall, these findings suggest that designing XR feedback involves considering both what a cue communicates and how its presentation shapes the experience of receiving it in front of others.
The same conversational state could be expressed through different metaphors, agents, or environmental cues. Future work could benefit from examining how different feedback forms affect both perceived social acceptability and actual conversational outcomes in live, multi-user XR settings.

\subsection{Calibrating Humor, Indirectness, and Attention Redirection}

\textbf{Humor Should Soften Feedback Without Trivializing the Situation.} Previous research has identified humor as a remedial strategy for reducing social embarrassment \cite{keltner1997embarrassment,fink1977humorous,clegg2012stranger,press2023humorous}. Consistent with prior work, our findings show that humor may provide group members with a socially acceptable way to acknowledge the corrective message. However, participants also mentioned that excessive humor could undermine the seriousness of the situation (Section \ref{sec:7.3.1}). This concern aligns with prior work about understanding the appropriateness of humor in group collaboration, which demonstrates that humor during serious conversation or goal-directed talk like a debate, lesson or business meeting could be perceived as intrusive, interruptive, or a waste of time \cite{norrick2008humor,vivona2014laugh}. Future work should adopt humor with care and only after considering the conversation context. Subtle playfulness may be sufficient for routine participation reminders \cite{banas2011review,bonaiuto2003arguing}, whereas explicitly humorous behavior may be excessive and distracting during serious contexts \cite{norrick2008humor,vivona2014laugh}.

\textbf{Decoupling Feedback from Individuals to Reduce Social Exposure.}
Prior work on the role of explicit social attention in amplifying embarrassment \cite{sabini2000embarrassed,clegg2012stranger,gilovich2000spotlight} informed our Strategy 2: Reduce social attention. Participants, however, noted that a new object could still expose a person if other members could easily infer whom it targeted. This finding adds an important nuance to the strategy: reducing social attention may require decoupling the feedback from a specific individual, rather than merely introducing a new element to the shared space. 
Building on participants’ proposals for environmental cues and agents acting as social facilitators (Section \ref{sec:7.3.2}), future work could examine whether group-level environmental changes or agent behaviors can redirect attention without singling out an individual, while keeping the intended feedback understandable.

\textbf{Indirectness should preserve face without obscuring the intended action.}
Previous HCI research has found subtle directives to be more socially acceptable than overt text-based instructions \cite{schiavo2014overt}, while linguistic research suggests that indirect expressions can protect people's public self-image \cite{crespo2005euphemistic,hadipragmatic}. Our findings similarly indicate that indirect metaphors and agent behaviors are perceived to reduce the confrontational quality of public feedback. However, indirectness introduced a tension between social comfort and interpretability: if the mapping between a cue and the expected behavioral adjustment is too weak, users may be uncertain about whether the feedback applies to them or how they should respond. Research on indirect communication further suggests that interpreting such cues often depends on shared cultural knowledge and communicative conventions \cite{joyce2012impact}. Future work should carefully calibrate the degree of indirectness. One possible way to reduce confusion is to familiarize users with the meanings of feedback cues in advance, for example through onboarding, a shared legend, or optional explanations.

\subsection{From Fixed to Flexible Feedback Delivery Mechanisms}
Our findings suggest that appropriate feedback delivery depends on the social context of the conversation. Participants considered public feedback more acceptable for positive reinforcement and in groups with greater familiarity and trust, while preferring private feedback for more sensitive situations, such as low participation associated with stress or lack of confidence (Section \ref{sec:7.1}). These preferences suggest a tension between supporting shared awareness and protecting individual comfort. Making feedback visible to the group may help coordinate participation, but it may also expose a person whose behavior calls for a more sensitive response.
Participants’ suggestions for gradual delivery and personalized feedback forms offer possible ways to accommodate these different needs (Section \ref{sec:7.3.4}). This aligns with prior work on customizable and context-relevant feedback \cite{samrose2018coco,samrose2021meetingcoach}, as well as user control over how non-verbal cues are shared \cite{vazquez2025trusting}. Future work could explore how feedback visibility, form, timing, and intensity can adapt to conversational contexts while allowing users to configure these decisions. Such systems should consider how to support group coordination while respecting recipients’ preferences about public exposure.

\section{Limitations and Future Work}

One limitation is that our study assessed participants' perceptions of the 3D feedback probes rather than during actual group conversations. Although the survey and follow-up interviews provided valuable insights into our work, participants evaluated visual sketches combined with text descriptions rather than experiencing these probes in a real-time interactive setting. Moreover, the study focused on the perspective of the feedback receiver and did not examine how other group members might perceive feedback directed at someone else. Future work could evaluate both receiver and spectator perspectives and investigate how these feedback designs influence participants' real-time conversational behavior and their perceptions in XR environments.

Second, as our study focused on a relatively limited set of scenarios and design probes, the generalizability of the results may be limited by the study context. We examined three conversational scenarios and a small number of feedback styles, which does not cover all the possible situations and feedback designs that may arise in small-group conversations. Additionally, co-located and distributed groups may respond differently to the feedback design due to differences in social presence and communication cues. Future work could extend this work by exploring a wider range of conversational scenarios, feedback styles, and interaction contexts.

Third, racial and ethnic diversity does not fully capture cultural factors relevant to the social acceptance of conversational feedback. We did not systematically examine participants’ country of residence/origin and cultural background, which limits the cross-cultural generalizability of findings. Future work should investigate how these factors shape feedback preferences and acceptance.

Finally, there was s time gap of approximately one month between the survey and interviews, and some participants developed different views on the relevant design in the later session. While this provided more nuanced insights for our work, it also suggests that participants’ perceptions were not entirely stable across different sessions. Future work could shorten this time gap or more systematically explore how participants' perceptions evolve over long-term use.

\section{Conclusion}

In this paper, we investigated how public feedback in XR-based small-group conversations can be designed to remain socially acceptable while preserving its effectiveness. Drawing on social science and HCI literature, we identified three design strategies for reducing embarrassment in social contexts: incorporating humor, reducing social attention, and using indirectness. Guided by these strategies and novel digital representations in XR, we developed a set of metaphor- and agent-based 3D feedback probes for three conversational scenarios: dominance, under-participation, and negative emotion. Through a two-phase study combining an online survey and follow-up interviews, our findings suggest that metaphor- and cat-based 3D feedback can mitigate some of the social discomfort associated with public feedback while maintaining perceived usefulness and fairness. Participants explained how design strategies made public feedback easier to accept, while also highlighting the trade-offs involved in applying them. Ultimately, this work contributes design implications for future socially acceptable 3D feedback in small-group conversations.

\bibliographystyle{ACM-Reference-Format}
\bibliography{references}

\clearpage
\appendix

\section{Survey questions}
\label{survey_q}

\begin{table}[H]
\centering
\label{tab:questionnaire}
\begin{tabular}{p{0.28\linewidth} p{0.65\linewidth}}
\toprule
\textbf{Metric} & \textbf{Statement} \\
\midrule
Usefulness &
It would help reduce dominant participation/encourage quieter speakers to contribute more/ease negative emotions during the conversation. \\

Fairness &
It would help us participate more openly and fairly. \\

Comfort &
I would feel comfortable receiving it during a conversation. \\

Embarrassment &
Receiving this feedback would make me feel embarrassed. \\

Distraction &
I would be distracted by it. \\

Incorporate humor &
It would be playful and easier to accept. \\

Reduce social attention &
It would avoid putting me at the center of attention. \\

Use indirectness &
It would be indirect and help me save my face. \\
\bottomrule
\end{tabular}
\caption{Questionnaire administered after viewing each feedback design probe. All items were rated on a 5-point Likert scale (1 = Strongly disagree, 5 = Strongly agree).}
\end{table}

\section{Open-ended responses}
\label{survey_op}

\begin{table}[H]
  \centering
  \caption{Participants’ perceived advantages, concerns, and suggested
  improvements regarding feedback probes. The numbers in parentheses
  indicate how many participants mentioned each statement.}
  \label{tab:feedback-perceptions}

  \small
  \renewcommand{\arraystretch}{1.3}
  \resizebox{\linewidth}{!}{%
    \begin{tabular}{@{}p{1.8cm}p{3.6cm}p{3.6cm}p{3.6cm}@{}}
      \toprule
      \textbf{Probes}
      & \textbf{Advantages}
      & \textbf{Concerns}
      & \textbf{Suggested Improvements}\\
      \midrule

      \textbf{\textit{Text} (Baseline)}
      & direct (8), clear (5), precise (2), easy to understand (4),
        efficient (1)
      & easy to get embarrassed (2), not warm (1), devoid of emotion (1),
        very serious and less fun (1), too straightforward (2),
        can come across as blunt (1), judgmental (1)
      & \multirow[t]{3}{3.6cm}{%
          \textit{Provide flexibility.}
          Allow users to choose their preferred feedback probes (2).
          The receiving person may have preferences on what works
          and what doesn't (1).
          Make the 3D assets change dynamically depending on the
          situation, or even make them interactive (1).
          \par\vspace{1em}
          \textit{Provide multimodal cues.}
          Voice reminder (1).
          Incorporate haptic pulses or spatial audio cues to ensure
          the feedback is accessible to everyone (1).
        }\\

      \cmidrule(lr){1-3}

      \textbf{\textit{Metaphor}}
      & engaging (1), easy to accept (1), less intrusive (3),
        reduce hostility (1), not too straightforward (1),
        less confrontational (2), without lowering self esteem (1)
      & not informative enough (1), indirect (3), ambiguous (4),
        sometimes could not receive the hints (1),
        a bit hard or trickier to understand (2),
        might not be easily noticeable (1)
      & \\

      \cmidrule(lr){1-3}

      \textbf{\textit{Cat}}
      & playful and funny (3), hilarious (1), amusing (1),
        attractive (1), interesting (1), ease awkwardness (1),
        gentle (1), emotionally supportive (3), friendly (1),
        low-stress (1), non-threatening (1)
      & indirect (5), lack the directness needed in more serious
        situations (1), sometimes could not receive the hints (1),
        distracting (1)
      & \\

      \bottomrule
    \end{tabular}%
  }
\end{table}

\section{Follow-up interview questions}
\label{interview_q}

\subsection{Public vs. private feedback}

1. [Interpret open-ended Responses for public feedback] In your responses, you mentioned that public feedback would \underline{\hspace{1cm}}(open-ended responses in survey)\underline{\hspace{1cm}}.
\begin{itemize}
    \item Can you tell me more about your perceptions?
    \item What other advantages/disadvantages do you think public feedback has?
    \item What do you think public feedback would affect shared awareness or coordination in the group? (Ask if not discussed in previous questions)
    \item In what kinds of situations do you think public feedback would be helpful? And in what situations would public feedback not work?
    \end{itemize}

2. Interpret open-ended Responses for private feedback: In your responses, you mentioned that private feedback would \underline{\hspace{1cm}}(open-ended responses in survey)\underline{\hspace{1cm}}.
\begin{itemize}
    \item Can you tell me more about your perceptions?
    \item What other advantages/disadvantages do you think private feedback has?
    \item What do you think private feedback would affect the social comfort of seeing the feedback? Some people suggested that private feedback may be easier to ignore. What do you think? (Ask if not discussed in previous questions)
    \item In what kinds of situations do you think private feedback would be helpful? And in what situations would private feedback not work?
\end{itemize}

\subsection{Under public condition, response to Text, Metaphor and Cat feedback designs}

1. [Interpret Ratings for usefulness] I noticed that you rated that cat feedback is more useful than other feedback (or other feedback is more useful).
\begin{itemize}
    \item Can you share your thoughts? Why do you think cat feedback is more useful?
    \item Why is text style less useful? Why is metaphor style less useful?
    \item {[Scenario]} Does that differ in different scenarios?
    \item {[Are they understanding usefulness differently]} When you evaluated perceived usefulness, what were you imagining would happen in the group?
\end{itemize}

2. [Interpret Ratings for fairness] I noticed that you rated that cat feedback would bring more fairness than other feedback (or that other styles would bring more fairness).
\begin{itemize}
    \item Can you share your thoughts? Why do you think cat feedback brings more fairness?
    \item Why does the text style bring less fairness? Why does metaphor style bring less fairness?
    \item {[Scenario]} Does that differ in different scenarios?
    \item {[Are they understanding fairness differently]} When you evaluated perceived fairness, what were you imagining would happen in the group?
\end{itemize}

3. [Interpret Ratings for distracting] I noticed that you rated that metaphor feedback is more distracting than other feedback (or that other feedback is more distracting).
\begin{itemize}
    \item Why: Can you share your thoughts?
    \item Why is the text style less distracting? Why is cat style less distracting?
    \item {[Scenario]} Does that differ in different scenarios?
\end{itemize}

4. [Interpret Ratings for comfort and embarrassment] I noticed that you rated that cat feedback makes you feel more comfortable and less embarrassed (or that other styles make you more comfortable).
\begin{itemize}
    \item Can you share your thoughts? Why do you think cat feedback makes you feel more comfortable and less embarrassed?
    \item {[Interpret strategy]} You rated higher for humor for cat feedback (or other DGs). Do you think it reduces embarrassment because of humor? I wonder if you think that might influence embarrassment. Can you tell us more?
    \item Ask about indirectness and reduced social attention (feel less “called out”)
    \item Why is the text style less comfortable?
    \item Why is metaphor style less comfortable? Ask about their ratings for humor, indirectness, and reduced social attention
    \item {[Scenario]} Does that differ in different scenarios?
\end{itemize}

5. Please rank the following six feedback types in order of your overall preference, from your most to least preferred. Briefly explain your ranking.
\begin{itemize}
    \item Cue the three scenarios
\end{itemize}

6. [Interpret open-ended Responses for cat/metaphor/text feedback] In your responses, you mentioned that cat/metaphor/text feedback would \underline{\hspace{1cm}}(open-ended responses in survey)\underline{\hspace{1cm}}. 
\begin{itemize}
    \item Can you tell me more about your perceptions?
    \item In what kinds of situations do you think cat feedback would be helpful? And in what situations would cat feedback not work?
\end{itemize}

\subsection{Additional design suggestions}

1. [improvements for each design] Are there other metaphors, characters, or visual styles that might work better?
\begin{itemize}
    \item e.g., cat behaviors
\end{itemize}

2. [other designs for each strategy] Beyond metaphor and cat styles, are there other ways to implement humor to reduce embarrassment?
\begin{itemize}
    \item Any other ways to implement reduced social attention?
    \item Any other ways to implement indirectness?
\end{itemize}

3. [other strategies] Beyond the designs shown, are there any other strategies that could reduce embarrassment caused by public feedback? 
\begin{itemize}
    \item What kind of feedback can be implemented to realize this strategy?
    \item If they cannot think of any, ask in real life, how do you reduce embarrassment in small group conversations?
\end{itemize}

\end{document}